\documentclass[aps,prl,twocolumn,showpacs,superscriptaddress,groupedaddress]{revtex4}

\usepackage{graphicx}
\usepackage{verbatim}
\usepackage{float}
\usepackage{sidecap}
\usepackage{dcolumn}
\usepackage{bm}
\usepackage{amssymb}
\usepackage{amsmath}	
\usepackage{color}
\usepackage{multirow}
\usepackage{lipsum}

\begin{document}

\title{Donor hyperfine Stark shift and the role of central-cell corrections in tight-binding theory}

\author{Muhammad Usman} \email{usman@alumni.purdue.edu} \affiliation{Center for Quantum Computation and Communication Technology, School of Physics, The University of Melbourne, Parkville, 3010, VIC, Australia.} 
\author{Rajib Rahman} \affiliation{Network for Computational Nanotechnology, School of Electrical and Computer Engineering, Purdue University, West Lafayette, 47906, Indiana, USA.} 
\author{Joe Salfi} \affiliation{Center for Quantum Computation and Communication Technology, School of Physics, The University of New South Wales, Sydney, NSW, 2052, Australia}
\author{Juanita Bocquel} \affiliation{Center for Quantum Computation and Communication Technology, School of Physics, The University of New South Wales, Sydney, NSW, 2052, Australia}
\author{Benoit Voisin} \affiliation{Center for Quantum Computation and Communication Technology, School of Physics, The University of New South Wales, Sydney, NSW, 2052, Australia}
\author{Sven Rogge} \affiliation{Center for Quantum Computation and Communication Technology, School of Physics, The University of New South Wales, Sydney, NSW, 2052, Australia}
\author{Gerhard Klimeck} \affiliation{Network for Computational Nanotechnology, School of Electrical and Computer Engineering, Purdue University, West Lafayette, 47906, Indiana, USA.} 
\author{Lloyd L. C. Hollenberg} \email{lloydch@unimelb.edu.au} \affiliation{Center for Quantum Computation and Communication Technology, School of Physics, The University of Melbourne, Parkville, 3010, VIC, Australia.} 

\vskip 0.25cm

\date{\today}

\begin{abstract}
Atomistic tight-binding (TB) simulations are performed to calculate the Stark shift of the hyperfine coupling for a single Arsenic (As) donor in Silicon (Si). The role of the central-cell correction is studied by implementing both the static and the non-static dielectric screenings of the donor potential, and by including the effect of the lattice strain close to the donor site. The dielectric screening of the donor potential tunes the value of the quadratic Stark shift parameter ($\eta_2$) from -1.3 $\times$ 10$^{-3} \mu$m$^2$/V$^2$ for the static dielectric screening to -1.72 $\times$ 10$^{-3} \mu$m$^2$/V$^2$ for the non-static dielectric screening. The effect of lattice strain, implemented by a 3.2\% change in the As-Si nearest-neighbour bond length, further shifts the value of $\eta_2$ to -1.87 $\times$ 10$^{-3} \mu$m$^2$/V$^2$, resulting in an excellent agreement of theory with the experimentally measured value of -1.9 $\pm$ 0.2 $\times$ 10$^{-3} \mu$m$^2$/V$^2$. Based on our direct comparison of the calculations with the experiment, we conclude that the previously ignored non-static dielectric screening of the donor potential and the lattice strain significantly influence the donor wave function charge density and thereby leads to a better agreement with the available experimental data sets.                           
\end{abstract}

\maketitle

\section{Introduction}

Since the Kane proposal for quantum computing using donor spins in silicon~\cite{Kane_Nature_1998}, there has been considerable progress towards the realisation of spin-qubit architectures~\cite{Zwanenburg_RMP_2013, Lloyd_PRB_2006, Hill_PRB_2005, Sousa_PRB_2009}. Notable results include single-atom fabricated devices~\cite{Fuechsle_NN_2012, Watson_NL_2014, Weber_Science_2012} and control and measurement of individual donor electron and nuclear spins~\cite{Tyryshkin_Nat_Mat_2012, Saeedi_Science_2013,  Morello_Nature_2010, Pla_Nature_2013}. However in building a scalable donor-based quantum computer, an important aspect is understanding and controlling the Stark shift of the donor hyperfine levels.

Accurate theoretical modelling of the donor hyperfine coupling is a challenging problem. First it requires proper incorporation of the valley-orbit (VO) interaction which has been established as a critical parameter to accurately match the experimentally observed energies of the ground state (A$_1$-symmtery) and the excited states (T$_2$ and E symmetries)~\cite{Wellard_Hollenberg_PRB_2005}. Secondly it is essential to perform the calculation of the ground state donor wave function with high precision through proper implementation of the central-cell effects (short-range potential) and the dielectric screening of long-range Coulomb potential. 

Earlier studies based on the Kohn and Luttinger's single-valley effective-mass theory (SV-EMT)~\cite{Kohn_PR_1955} ignored the VO interaction and therefore could not match with the experimental binding energy of the ground state (A$_1$); since then several studies have been performed with incremental improvements in the model. Pantelides and Sah~\cite{Pantelides_Sah_PRB_1974} pointed out that the concept of the central-cell correction is ill defined in SV-EMT and therefore it fails to capture the chemical shift and the splitting of the experimentally observed donor ground state energies which primarily arise from the intervally mixing. Based on this, they presented a multi-valley effective-mass theory (MV-EMT) by explicitly including the central-cell correction along with a non-static dielectric screening of the Coulomb potential representing the donor. Similar EMT based formalisms have been widely applied by various studies later on to investigate the physics of shallow donors~\cite{Saraiva_1, King_1, Pica_1}. 

Overhof and Gerstmann~\cite{Overhof_PRL_2004} applied density-functional theory to successfully calculate the hyperfine ferquency of shallow donors in Si. While their calculations were in excellent agreement with experiment (zero fields), the ab-initio description of the donor wave function was by definition limited to only a few atoms around the donor site. Also they ignored the long-range tail of the Coulomb potential and therefore were unable to match the donor binding energies from their approach. More recently, a much more detailed theoretical calculation was performed by Wellard and Hollenberg~\cite{Wellard_Hollenberg_PRB_2005} based on band-minimum basis (BMB) approach. In their study, by using a core-correcting potential screened by non-static ($k$-dependent) dielectric function, they were able to demostrate excellent agreement with the experimentally measured ground state energy for the P donor in Si (45.5 meV). However the excited state energies remained few meVs off from the experiemental values. Nevertheless, their study clearly highlighted the critical role of the central-cell corrections in theoretical modeling of shallow donors in Si which drastically modifies the charge density of donor wave function and therefore tune the hyperfine coupling and its Stark shift parameters.

Atomistic tight-binding (TB) approach, historically used for the deep level impurities~\cite{Vogl_SSP_1981}, has been shown to work remarkably well for the shallow level impurities in Si~\cite{Martins_PRB_2004, Rahman_PRL_2007, Weber_Science_2012}. The TB method offers several advantages over EMT and DFT based methodologies, including the capability of inherently incorporating the VO intermixings, calculations over very large supercells (containing several million atoms in the simulation domain) and therefore providing much more detailed description of the donor wave funtion, an easy incorporation of externally applied electric field effects in Hamiltonian, and explicit representation of the short-range and the long-range donor potentials, etc. Martin \textit{et al.}~\cite{Martins_PRB_2004} applied second neares-neighbor sp$^3$s$^*$ TB model to study the effects of an applied electric field on P donor wave functions. Later Rahman \textit{et al.}~\cite{Rahman_PRL_2007} applied a much more sophisticated sp$^3$d$^5$s$^3$ TB Hamiltonian to P donors in Si and bench-marked Stark shift of the donor hyperfine coupling against the BMB calculations. The two theoretical models were found to be in remarkable agreement with each other for P donors, and also exhibited very good agreement with the Stark shift measurement data for the Sb donor in Si~\cite{Bradbury_PRL_2006}; however a direct comparison of the hyperfine Stark shift with experiment was not possible due to the unavailability of any experimental data for the P and As donors. The models were also based on minimal central-cell correction, implemented in terms of short-range correction of donor potential at the donor site and a static dielectric screening of the long-range donor potential tail. 

The previously reported experimental data for the hyperfine coupling of the shallow donors (P, As, Sb etc.) in Si~\cite{Feher_PR_1959}, and more recent experimental measurements~\cite{Lo_arxiv_2014} of the hyperfine Stark shift for As donor in Si provide excellent opportunities to directly bench-mark TB theory against the experiment data sets. This work for the first time evaluates the role of central-cell corrections in atomistic TB theory through a direct comparison against the experimental data of the hyperfine interaction for a single As donor in Si. 

The central-cell corrections in the tight-binding model considered here are implemented by:
\begin{itemize}

\item[1)] Short-range correction of the donor potential: donor potential is truncated at U$_0$ at the donor site.
\item[2)] Dielectric screening of the long-range tail of the donor potential: static vs. non-static dielectric screenings.
\item[3)] Lattice strain around the donor site: changes in the nearest-neighbor bond lengths.

\end{itemize}
\noindent
We systematically study the critical significance of the central-cell corrections by including the effect of each central-cell component one-by-one and evaluating its role on the hyperfine coupling and its Stark shift parameter. In each case, we first adjust U$_0$ to match the experimentally measured donor binding energies for the ground and excited states within 1 meV accuracy~\cite{Ahmed_Enc_2009}. We then compute charge density at the nuclear site and its character under the influence of an external electrical field. Our calculations demonstrate that the previously ignored central-cell components, non-static dielectric screening of donor potential and lattice strain, produce significant impact on the donor herperfine Stark shift and therefore lead to match the experimental data with an unprecedent accuracy. Such high precision bench-marking of the theory against the experimental data would be useful in accomplishing high precision control over donor wave functions required in quantum computing.  

\begin{SCfigure*}
\includegraphics[scale=0.55]{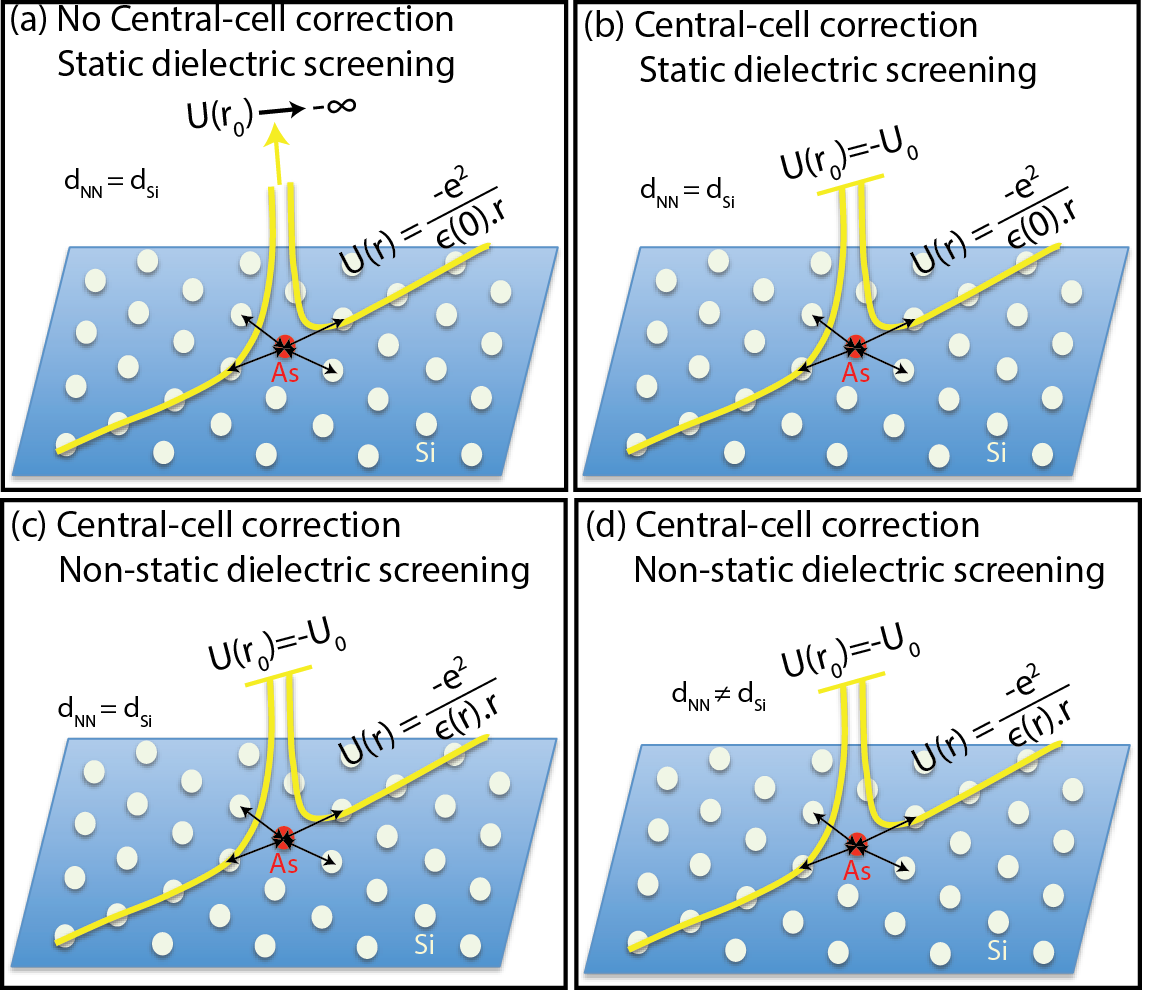}
\caption{The schematic diagram illustrating the flow of our study of the central-cell corrections and the donor potential screening: (a) Simulations are performed with static dielectric screening ($\epsilon(0)$) of the donor potential, no changes in the neareast-neighbor (NN) As-Si bond-length (d$_{NN}$) which is kept same as unstrained Si bond length (d$_{Si}$), and no inclusion of the central-cell corrections. This mimicks a simple case of effective-mass type calculation with no central-cell correction included. (b) Minimal central-cell correction is implemented through U$_0$ at the donor site and the donor potential is screened with static dielectric constant ($\epsilon(0)$). No changes in the NN bond-lengths are included. (c) The effect of non-static dielectric screening ($\epsilon(r)$) is included in the simulations. (d) All of the three central-cell effects (U$_0$, $\epsilon(r)$, and d$_{NN} \neq$ d$_{Si}$) are included in the simulations.}
\label{fig:Fig1}
\end{SCfigure*}
\vspace{1mm}

\section{Methodology}

The atomistic simulations are performed using NanoElectronic MOdeling tool NEMO-3D~\cite{Klimeck_1, Klimeck_2}, which has previously shown to quantitatively match the experimental data sets for a variety of nanostructures and nanomaterials, such as shallow donors in Si ~\cite{Rahman_PRL_2007, Weber_Science_2012}, III-V alloys~\cite{Usman_1, Usman_2} and quantum dots~\cite{Usman_3, Usman_4, Usman_5}, SiGe quantum wells~\cite{Neerav_1}, etc. The sp$^3$d$^5$s$^*$ tight-binding parameters for Si material are obtained from Boykin \textit{et al}.~\cite{Boykin_PRB_2004}, that have been optimised to accurately reproduce the Si bulk band structure. The As donor is represented by a screened Coulomb potential truncated to U($r_0$)=U$_0$ at the donor site, $r_0$. Here, U$_0$ is an adjustable parameter that represents the central-cell correction at the donor site and has been designed to accurately match the ground state binding energy (A$_1$ = 53.8 meV) of the As donor as measured in the experiment. The size of the simulation domain (Si box around the As donor) is chosen as 32 nm $\times$ 65 nm $\times$ 32 nm, consisting of roughly 3.45 million atoms, with closed boundary conditions in all three dimensions. The surface atoms are passivated by our published method~\cite{Lee_PRB_2004} to avoid any spurious states in the energy range of interest. The multi-million atom real-space Hamiltonian is solved by a prallel Lanczos algorithm to calculate donor single-particle energies and wave functions. 

For the study of the effects of the lattic relaxation, the influence of the changed nearest-neighbor bond lengths on the tight-binding Hamiltonian is computed by a generalization of the Harrison's scalling law~\cite{Boykin_PRB_2004}. In this formulation, the interatomic interaction energies are taken to vary with the bond length $d$ as $(\dfrac{d_0}{d})^\eta$, where d$_0$ is the unrelaxed Si bond length and $\eta$ is a scalling parameter whose magnitude depends on the type of the interaction being considered and is fitted to obtain hydrostatic deformation potentials.

The hyperfine coupling parameter A(0) is directly proportional to the squared magnitude of the ground state wave function at the donor nuclear site, $|\psi(r_0)|^2$~\cite{Rahman_PRL_2007} and its value is experimentally measured~\cite{Feher_PR_1959} as 1.73 $\times$ 10$^{30}$  m$^{-3}$ for As donor. It is therefore important to theoretically compute the value of $|\psi(r_0)|^2$  at the donor site and compare it with the experimental value. It should be pointed out that in our empirical tight-binding model, the Hamiltonian matrix elements comprising the onsite and nearest-neighbor interactions are optimized numerically to fit the bulk band structure of the host Si material without explicit knowledge of the underlying atomic orbitals. Therefore it is fundamentally not possible to quantitatively determine the value of the hyerperfine coupling A(0) as is possible from the ab-initio type calculations~\cite{Overhof_PRL_2004}. Nevertheless, we apply the methodology published by Lee \textit{et al.}~\cite{Lee_JAP_2005} to estimate the value of $|\psi(r_0)|^2$ from our model, where we have used the value of bulk Si conduction electron at the nuclear site as $\approx$ 9.07 $\times$ 10$^{24}$ cm$^{-3}$~\cite{Shulman_PR_1956, Lucy_PRB_2011} and the value of the atomic orbital ratio $\phi_{s^*}(0)/\phi_{s}(0)$ computed to be 0.058 from the assumption of the hydrogen-like atomic orbitals with an effective nuclear charge~\cite{Clementi_JCP_1963}. We believe that this provides a good qualitative comparison of A(0) $ \propto |\psi(r_0)|^2$ from our model with the experimental value, and along with the quantitative match of the donor binding energies (A$_1$, T$_2$, and E) and the Stark shift of hyerfine ($\eta_2$), serve as a benchmark to evaluate the role of the central-cell corrections in the tight-binding theory.  

We calculate the Stark shift of the hyperfine interaction as follows~\cite{Rahman_PRL_2007}: the potential due to the electrical field is added in the diagonal of the tight-binding Hamiltonian which distorts the donor wave function and pulls it away from the donor site reducing the field dependent hyperfine coupling, A($\overrightarrow{E}$); the hyperfine coupling A($\overrightarrow{E}$) is directly proportional to $|\psi ( \overrightarrow{E}, r_0 )|^2$, where $r_0$ is the location of donor. The change in A($\overrightarrow{E}$) is parametrized as:

\begin{equation}
	\label{eq:hyperfine_coupling}
	\Delta A \left( \overrightarrow{E} \right) = A \left( 0 \right) \left( \eta_2 E^2 + \eta_1 E \right)
\end{equation}

\noindent 
Here $\eta_2$ and $\eta_1$ are the quadratic and linear components of the Stark shift of the hyperfine interaction, respectively. For deeply burried donors (with donor depths typically greater than about 15 nm), the linear component of the Stark shift becomes negligible~\cite{Rahman_PRL_2007}. Therefore we do not provide values of $\eta_1$ in the remainder of this paper which are about two to three orders of magnitude smaller than the values of $\eta_2$. 

\section{Screening of donor potential by static dielectric constant}

In the first set of simulations, we apply no central-cell correction (U$_0 \rightarrow - \infty$) and the long-range part of the donor potential is Coulomb potential screened by static dielectric constant ($\epsilon(0)$) as given by Eq.~\ref{eq:Static_donor_potential}: 

\begin{equation}
	\label{eq:Static_donor_potential}
	U \left( r \right) = \frac{-e^2}{ \epsilon \left( 0 \right) r} 
\end{equation}

\noindent
where $\epsilon(0)$ = 11.9 is the static dielectric constant of Si and $e$ is the charge on electron. This case is illustrated by schematic of Fig.~\ref{fig:Fig1} (a). Such setup leads to a six-fold degenerate set of donor states at a binding energy of $\approx$ 29.6 meV, as would be expected from a simple effective-mass type approximation. The donor wave function density at nuclear site is 3.57$\times$10$^{22}$ m$^{-3}$ which is seven orders of magnitude smaller than the experimental value. This clearly highlights the critical role of the central-cell correction at the donor site to accurately capture the splitting of the donor ground state binding energies and the wave function density at the nuclear site as measured in the experiment.

\begin{figure}
\includegraphics[scale=0.45]{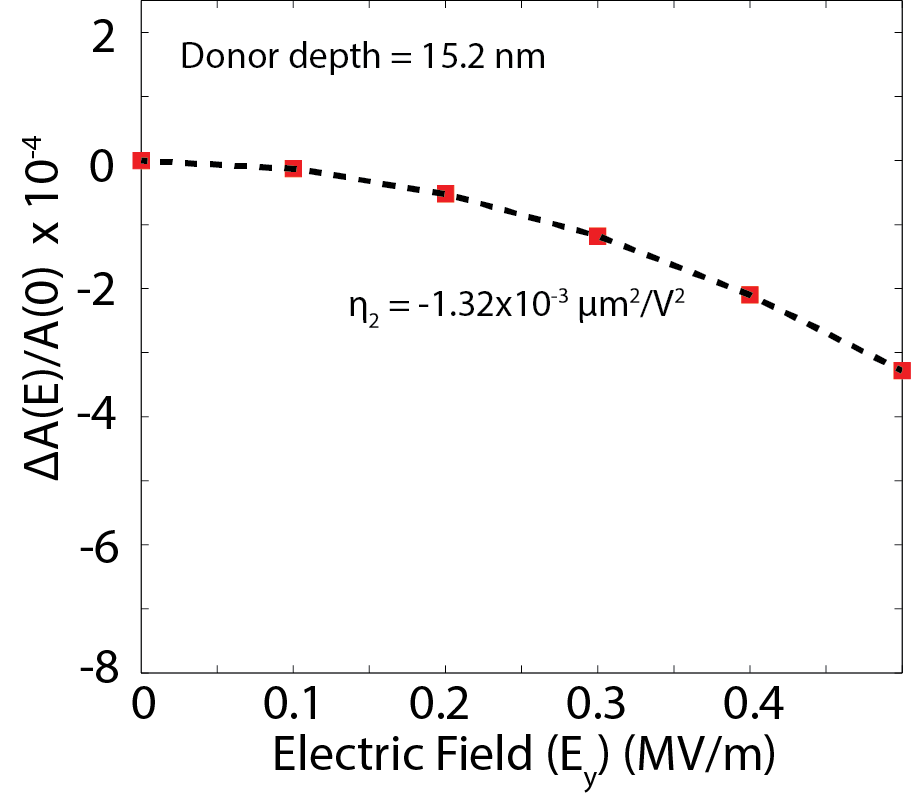}
\caption{Electric field response of hyperfine coupling is plotted for bulk As donor for static dielectric screening of donor potential as given by Eq.~\ref{eq:Static_donor_potential}. The data points are computed directly from the TB simulations and the line plots are fittings of Eq.~\ref{eq:hyperfine_coupling}. }
\label{fig:Fig2}
\end{figure}
\vspace{1mm}

\begin{table*}[ht!]
\vspace{0ex}
\caption{\label{tab:table1} Comparison of the calculated and the experimentally measured values for the As donor in Si. The experimental values of the donor binding energies (A$_1$, T$_2$, and E) are taken form Ref.~\onlinecite{Ramdas_RPP_1981}, the value of $|\psi(r_0)|^2$ is taken from Ref.~\onlinecite{Feher_PR_1959}, and the measured value of the quadratic hyperfine Stark shift $\eta_2$ is taken from Ref.~\onlinecite{Lo_arxiv_2014}. The TB calculations are based on the static dielectric screening of the donor potential (Eq.~\ref{eq:Static_donor_potential}). }
\small{\begin{tabular}
{@{\hspace{0.5ex}}l@{\hspace{0.2cm}}c@{\hspace{0.2cm}}c@{\hspace{0.2cm}}c@{\hspace{0.2cm}}c@{\hspace{0.2cm}}c@{\hspace{0.2cm}}c@{\hspace{0.5cm}}c}
\\
\hline
\hline
                        & Central-cell (U$_0$) & Static & A$_1$  & T$_2$  & E & $|\psi(r_0)|^2$ & $\eta_2$   \\
		      &  (eV) & Screening & (meV)  & (meV) & (meV) & (m$^{-3}$) & ($\times$ 10$^{-3} \mu$m$^2$/V$^2$) \\
\hline
Experiment             & - & - & 53.8 & 32.7 & 31.3 & 1.73$\times$10$^{30}$ & -1.9 $\pm$ 0.2 \\
TB Theory    & 2.6342  & $\epsilon(0)$ & 53.1 & 32.0 & 30.6 & 4.05$\times$10$^{30}$ & -1.32 \\
\hline
\hline
\end{tabular}}
\end{table*}         

\begin{table*}[ht!]
\vspace{0ex}
\caption{\label{tab:table2} The fitting values of $\epsilon(0)$, A, $\alpha$, $\beta$, and $\gamma$ given by several studies are taken from Ref.~\onlinecite{Sarker_JPC_1977} and the references therein.}
\small{\begin{tabular}
{@{\hspace{0.6ex}}l@{\hspace{0.8cm}}c@{\hspace{0.8cm}}c@{\hspace{0.8cm}}c@{\hspace{0.8cm}}c@{\hspace{0.8cm}}c@{\hspace{0.8cm}}c}
\\
\hline
\hline
         Non-static Dielectric  & $\epsilon(0)$  & A  & $\alpha$ & $\beta$ & $\gamma$  \\
		      Screenings & - & - & (au) & (au) & (au) \\
\hline \hline

Pantelides \& Sah (P\&S)         & 11.4 & 1.1750 & 0.7572 & 0.3223 & 2.044 \\
Nara \& Morita (N\&M)              & 10.8 & 1.1750 & 0.7572 & 0.3223 & 2.044 \\
Walter \& Cohen (W\&C)           & 11.3 & 1.0000 & 0.9500 & 0.0000 & 2.044 \\
Richardson \& Vinsome (R\&V) & 10.8 & 0.8918 & 0.9743 & 0.1586 & 0.1586 \\
\hline
\hline
\end{tabular}}
\end{table*}

Next, we setup simulations according to the schematic of Fig.~\ref{fig:Fig1} (b), where we keep the donor potential U($\overrightarrow{r}$) as a Coulomb potential screened by static dielectric constant for Si as given by Eq.~\ref{eq:Static_donor_potential}. Previous tight binding based theoretical studies for the P donors~\cite{Rahman_PRL_2007, Martins_PRB_2004} and the As donors~\cite{Lansbergen_Nat_Phys_2008} has also used this type of donor potential. We now include central-cell correction at the donor site as a cut-off potential U$_0$ which is tuned to be 2.6342 eV to accurately reproduce the experimental ground state energy, A$_1$=53.1 meV. Further tuning of the onsite TB $d-$orbital energies~\cite{Ahmed_Enc_2009} allowed to match the experimental excited state energies (T$_2$ and E) as listed in table~\ref{tab:table1}. By applying this model, we compute the value of $|\psi(r_0)|^2$ at the donor nuclear site as 4.05 $\times$ 10$^{30}$, which comes out to be $\approx$ 2.34 times larger than the experimental value. We also compute the electric field response of the hyperfine coupling for the electric field variation from 0 to 0.5 MV/m as shown in the Fig.~\ref{fig:Fig2}. The quadratic hyperfine Stark shift parameter $\eta_{2}$ is then calculated from the fitting of the TB data by Eq.~\ref{eq:hyperfine_coupling} (details of the calculation methodology have been reported in Ref.~\onlinecite{Rahman_PRL_2007}) as -1.32 $\times$ 10$^{-3} \mu$m$^2$/V$^2$ compared to the recent experimental value of -1.9 $\pm$ 0.2 $\times$ 10$^{-3} \mu$m$^2$/V$^2$ for the bulk As donor in Si~\cite{Lo_arxiv_2014}. Table~\ref{tab:table2} provides an overall summary of results for the static dielectric screening of the donor potential. This shows that even with the static dielectric screening of the donor potential, the central-cell correction part provides a reasonably good description of the donor physics. 

Since the central-cell effects are implemented through an adjustable parameter U$_0$ at the donor site, we attempt to quantify its effect on the $\eta_2$ by introducing a variation of $\pm$100 meV in its value. Increasing U$_0$ by 100 meV increases the ground state binding energy to 55.6 meV and the value of $\eta_2$ decreases to -1.089 $\times$ 10$^{-3} \mu$m$^2$/V$^2$. On the other hand, decreasing U$_0$ by 100 meV decreases the ground state binding energy to 50.9 meV and the value of $\eta_2$ increases to -1.53 $\times$ 10$^{-3} \mu$m$^2$/V$^2$. This clearly demonstrates that to improve the match with the experimental value of $\eta_2$, the value of U$_0$ should be reduced; however this introduces a large error in the binding energy of the donor ground state which is clearly unacceptable. Therefore we conclude that the current TB model with central-cell parameter U$_0$ and the static dielectric screening of the donor potential provides, at the best, a value of -1.32 $\times$ 10$^{-3} \mu$m$^2$/V$^2$ for the Stark shift of the hyperfine coupling. In the next two sections, we include the effects of non-static dielectric screening of the donor potential and the effect of the nearest-neighbour bond length changes to further evaluate the performance of our TB model.

\section{Screening of donor potential by non-static dielectric function}

With the established tight-binding model as our test system, we start further investigation of the central-cell correction, in particular the screening of the donor potential in the vicinity of the As donor. Previous tight-binding calculations~\cite{Rahman_PRL_2007, Martins_PRB_2004} for the P donor in Si have been based on the static dielectric screening of the donor potential; however Wellard and Hollenberg~\cite{Wellard_Hollenberg_PRB_2005} have already demonstrated the critical importance of the non-static dielectric screening of the donor potential in their band minimum basis (BMB) calculations. By incorporating a non-static screening of the donor potential given in Ref.~\onlinecite{Pantelides_Sah_PRB_1974}, they computed an excellent agreement of the donor ground state binding energy with the experimental value. Furthermore, the effect of the non-static dielectric screening was in particular profound on the spatial distribution of the donor wave function around the donor site. Therefore it is critical to investigate the impact of non-static dielectric screening of the donor potential on the values of $|\psi(r_0)|^2$ and $\eta_2$ computed from the tight-binding model. In this section, we investigate this effect by incorporating various non-static dielectric screenings of the donor potentials as reported in the literature.  

\begin{figure}
\includegraphics[scale=0.45]{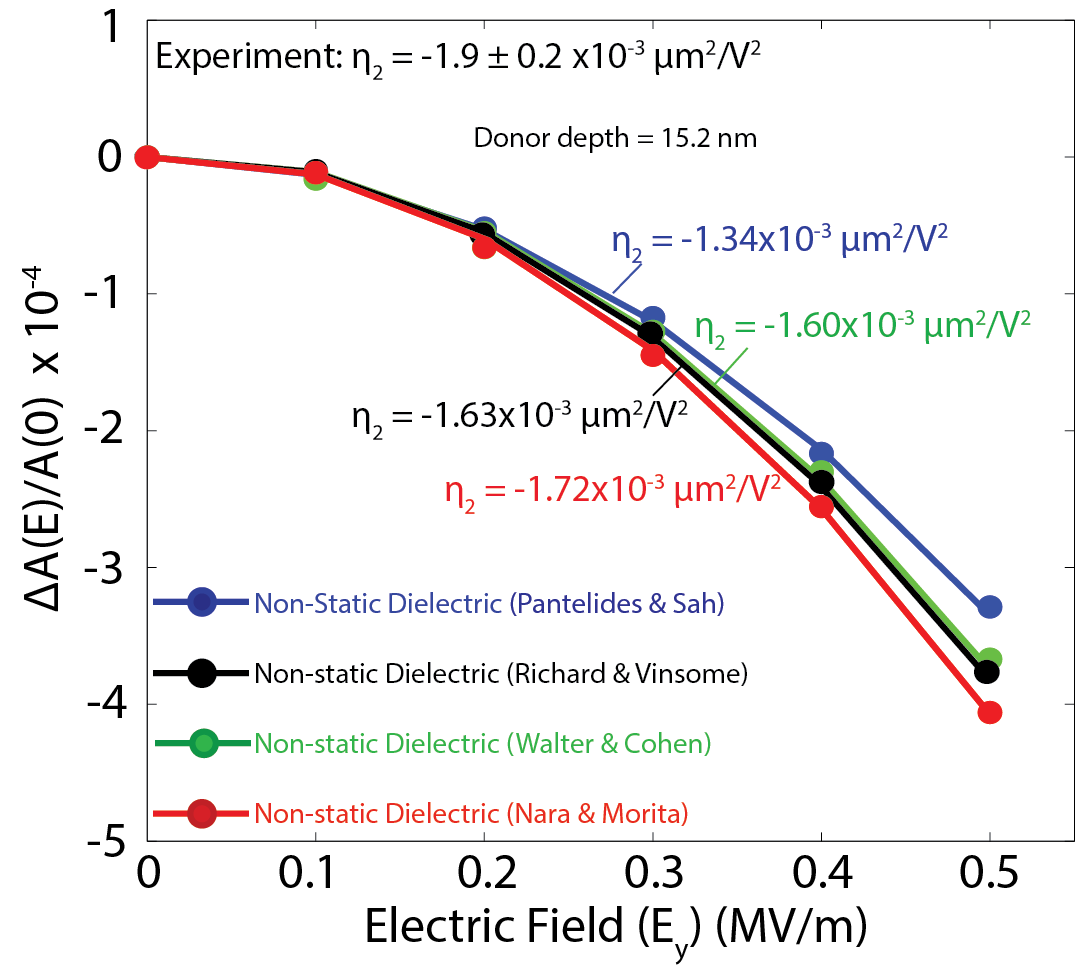}
\caption{Electric field response of hyperfine coupling is plotted for bulk As donor for various screenings of the donor potential. The data points are computed directly from the TB simulations and the line plots are fittings of Eq.~\ref{eq:hyperfine_coupling}. }
\label{fig:Fig3}
\end{figure}
\vspace{1mm}

The screening of the donor potential by a non-static dielectric constant has been a topic of extensive research, and a number of reliable calculations exist for $k$-dependent dielectric function, $\epsilon(k)$, for Si. The most commonly applied dielectric function is obtained by Nara~\cite{Nara_JPSJ_1965}: 

\begin{equation}
	\label{eq:Nonstatic_dielectric}
	\frac{1}{\epsilon(k)} = \frac{A^2 k^2}{k^2 + \alpha^2} + \frac{\left( 1-A \right) k^2}{k^2 + \beta^2} + \frac{1}{\epsilon \left( 0 \right) } \frac{\gamma^2}{k^2 + \gamma^2} 
\end{equation}

\noindent
where $A$, $\alpha$, $\beta$, and $\gamma$ are fitting constants and have been numerically fitted by various studies for Si (see table 2 for the fitting values reported by various authors). Based on this $k$-dependent dielectric constant, the new screened donor potential in the real space coordinate system is given by:

\begin{widetext}
\begin{equation}
	\label{eq:Nonstatic_donor_potential}
	U \left( r \right) = \frac{-e^2}{ \epsilon \left( 0 \right) r} \left( 1 + A \epsilon \left( 0 \right) \mathrm{e}^{- \alpha r} + \left( 1-A \right) \epsilon \left( 0 \right) \mathrm{e}^{- \beta r} - \mathrm{e}^{- \gamma r}  \right)
\end{equation}  
\end{widetext}

\begin{table*}[ht!]
\vspace{0ex}
\caption{\label{tab:table3} Comparison of the calculated and the experimentally measured parameters for As donor in Si. The experimental values of the donor binding energies (A$_1$, T$_2$, and E) are taken form Ref.~\onlinecite{Ramdas_RPP_1981}, the value of $|\psi(r_0)|^2$ is taken from Ref.~\onlinecite{Feher_PR_1959}, and the value of the quadratic hyperfine Stark shift $\eta_2$ is taken from Ref.~\onlinecite{Lo_arxiv_2014}. The TB calculations are based on the non-static dielectric screening of the donor potential (Eq.~\ref{eq:Nonstatic_donor_potential}). }
\small{\begin{tabular}
{@{\hspace{0.2ex}}l@{\hspace{0.2cm}}c@{\hspace{0.2cm}}c@{\hspace{0.2cm}}c@{\hspace{0.2cm}}c@{\hspace{0.2cm}}c@{\hspace{0.2cm}}c@{\hspace{0.2cm}}c}
\\
\hline
\hline
                        & Central-cell (U$_0$) & Non-static & A$_1$  & T$_2$  & E & $|\psi(r_0)|^2$ & $\eta_2$   \\
		      & (eV)  & Screening &  (meV)  & (meV) & (meV) & (m$^{-3}$) & ($\times$ 10$^{-3} \mu$m$^2$/V$^2$) \\
\hline
Experiment           & - & - & 53.8 & 32.7 & 31.3 & 1.73$\times$10$^{30}$ & -1.90 $\pm$ 0.2 \\
TB Theory            & 2.8842 & $\epsilon(k)$ (P\&S)      & 53.8 & 32.6 & 31.0 & 4.50$\times$10$^{30}$ & -1.34 \\
TB Theory            & 2.2842 & $\epsilon(k)$ (N\&M)           & 53.6  & 33.8 & 32.0 & 3.01$\times$10$^{30}$ & -1.72 \\    
TB Theory           & 2.0592 & $\epsilon(k)$ (W\&C)          & 53.5 & 33.2 & 31.8 & 3.36$\times$10$^{30}$ & -1.60 \\
TB Theory           & 2.6342 & $\epsilon(k)$ (R\&V)  & 53.9 & 34.2 & 32.7 & 2.65$\times$10$^{30}$ & -1.63 \\
\hline
\hline
\end{tabular}}
\end{table*} 

\begin{table*}[ht!]
\vspace{0ex}
\caption{\label{tab:table4} Comparison of the calculated and the experimentally measured parameters for As donor in Si. The experimental values of the donor binding energies (A$_1$, T$_2$, and E) are taken form Ref.~\onlinecite{Ramdas_RPP_1981}, the value of $|\psi(r_0)|^2$ is taken from Ref.~\onlinecite{Feher_PR_1959}, and the value of the quadratic hyperfine Stark shift $\eta_2$ is taken from Ref.~\onlinecite{Lo_arxiv_2014}. The TB calculations are based on the non-static dielectric screening of the donor potential (Eq.~\ref{eq:Nonstatic_donor_potential}). }
\small{\begin{tabular}
{@{\hspace{0.2ex}}l@{\hspace{0.2cm}}c@{\hspace{0.2cm}}c@{\hspace{0.2cm}}c@{\hspace{0.1cm}}c@{\hspace{0.1cm}}c@{\hspace{0.1cm}}c@{\hspace{0.2cm}}c@{\hspace{0.2cm}}c}
\\
\hline
\hline
                        & NN bond length (d$_{NN}$) & Central-cell (U$_0$) & Non-static & A$_1$  & T$_2$  & E & $|\psi(r_0)|^2$ & $\eta_2$   \\
		      & (nm) & (eV)  & Screening &  (meV)  & (meV) & (meV) & (m$^{-3}$) & ($\times$ 10$^{-3} \mu$m$^2$/V$^2$) \\
\hline
Experiment        & - & - & - & 53.8 & 32.7 & 31.3 & 1.73$\times$10$^{30}$ & -1.9 $\pm$ 0.2 \\
TB Theory           & 0.235 & 2.2842 & $\epsilon(k)$ (N\&M)           & 53.6  & 33.8 & 32 & 3.01$\times$10$^{30}$ & -1.72 \\    
TB Theory           & 0.2425 & 2.2842 & $\epsilon(k)$ (N\&M)          & 43.5 & 34.4 & 32.7 & 1.25$\times$10$^{30}$ & - \\
TB Theory           & 0.2425 & 3.1895 & $\epsilon(k)$ (N\&M)  & 53.5 & 33.87 & 31.8 & 2.96$\times$10$^{30}$ & -1.87 \\
\hline
\hline
\end{tabular}}
\end{table*} 

In our next set of simulations, we apply this donor potential and re-adjust the central-cell correction U$_0$ at the donor site to match the ground and excited state binding energies with the experimental values. The new values of U$_0$ and the corresponding values of the binding energies for A$_1$, E, and T$_2$ states are provided in the table~\ref{tab:table3} for the four non-static dielectric screenings of the donor potential under consideration in this study. After achieving this excellent agreement of the binding energies with the experimental values, we then compute the values of $|\psi(r_0)|^2$ and the electric field response of the hyperfine coupling as plotted in Fig.~\ref{fig:Fig3} for the various non-static dielectric screening potentials as listed in table~\ref{tab:table2}. The calculated values of $|\psi(r_0)|^2$ and $\eta_2$ are listed in table~\ref{tab:table3}. Overall, the non-static dielectric screening of the donor potential works remarkably well in the tight-binding theory and improves the match with the experimental values of $|\psi(r_0)|^2$ and $\eta_2$. For the non-static dielectric screening provided by Nara \& Morita, the agreement of the computed $\eta_2$ with the experimental value is within the range of experimental tolerance. We also find a direct relation of $|\psi(r_0)|^2$ with the central-cell correction parameter U$_0$. The smallest value of U$_0$ is for Richard \& Vinsome screening which results in the best match of $|\psi(r_0)|^2$ with the experimental value, different only by a factor of 1.5.  

\section{Effect of lattice relaxation}

In the calculations performed above so far, we have assumed the crystal lattice as perfect Si lattice where each atom including the As donor is connected to its four nearest neighbor (NN) atoms by unstrained bond lengths of 0.235 nm. In the past tight-binding studies of the donor hyperfine Stark shift~\cite{Rahman_PRL_2007, Martins_PRB_2004} the effect of lattice strain has been completely ignored based on the assumption that in the presence of the donor, the NN bond length only negligibly changes. However the recent ab-initio study~\cite{Overhof_PRL_2004} suggested a sizeable increase of 3.2\% in the NN bond length for the As donors in Si. Our fully atomistic description of the As donor in Si provides an excellent opportunity to investigate the effect of lattice strain. In our next set of simulations, we increase the bond length of the As donor and its four nearest Si neighbors by 3.2\%, thereby increasing it from the unstrained value of 0.235 nm to 0.2425 nm. For this study, we choose the non-static dielectric screening of the donor potential as described by Eq.~\ref{eq:Nonstatic_donor_potential} and the fitting parameters provided by Nara $\&$ Morita as given in the table~\ref{tab:table2}. This setup is schematically shown in Fig.~\ref{fig:Fig1} (d). Keeping the central-cell correction fixed at 2.2842 eV, we calculate a significant effect of the NN bond length change on the donor binding energies and the donor wave function confinement at the nuclear site. As evident from the third row of the table~\ref{tab:table4}, the donor ground state binding energy A$_1$ decreases by $\approx$ 10 meV and the value of $|\psi(r_0)|^2$ is decreased by a factor of $\approx$ 2.4 as a result of the lattice strain. 

Since the binding energies of the donor are adjusted in our model by central-cell correction (by varying U$_0$ and onsite TB energies), we perform further adjustments in U$_0$ by increasing its value to 3.1895 eV to re-establishe the match of the ground state binding energies with the experimental values. Based on this new model, we then recalculate the values of $|\psi(r_0)|^2$ and the Stark shift parameter $\eta_2$, and the corresponding values are provided in the last row of table\ref{tab:table2}. The lattice strain only slightly modifies the value of $|\psi(r_0)|^2$ at the donor site, howevere the quadratic Stark shift parameter $\eta_2$ is strongly affected and becomes -1.87 $\times$ 10$^{-3} \mu$m$^2$/V$^2$ which is in remarkable agreement with the exerimental value of -1.9 $\pm$ 0.2 $\times$ 10$^{-3} \mu$m$^2$/V$^2$. Further investigation is needed to establish the connection of the NN bond length change with the value of $|\psi(r_0)|^2$ which would be reported somewhere else.            

\section{Conclusions}

In conclusion, this work aims to evaluate and benchmark previously established tight-binding model with the recently measured experimental data of the quadratic Stark shift of the As donor hyperfine interaction. The study is systematically performed to investigate the central-cell correction effects in the tight-binding theory. We include central-cell corrections in terms of donor potential cut-off at the nuclear site, static vs. non-static dielectric screenings of the donor potential, and the effect of the lattice strain by changing the As-Si nearest-neighbor bond lengths. Overall our calculations exhibit that tight-binding theory captures the donor physics remarkably well by reproducing the donor binding energy spectra within 1 meV of the expereimentally measured values. When we include the effects of non-static dielectric screening of the donor potential and lattice strain, the computed value of the quadratic Stark shift parameter ($\eta_2$) is calculated to be -1.87 $\times$ 10$^{-3} \mu$m$^2$/V$^2$ which is in excellent agreement with the experimental value of -1.9 $\pm$ 0.2 $\times$ 10$^{-3} \mu$m$^2$/V$^2$. Such detailed bench-marking of theory against the experimental data would allow us to relaibly investigate the single and two donor electron wave functions, especially those relevant for implementing quantum information processing.            

\textbf{\textit{Acknowledgements:}} This work is funded by the ARC Center of Excellence for Quantum Computation and Communication Technology (CE1100001027), and in part by the U.S. Army Research Office (W911NF-08-1-0527).  Computational resources are acknowledged from National Science Foundation (NSF) funded Network for Computational Nanotechnology (NCN) through \url{http://nanohub.org}. NEMO 3D based open source tools are available at: \url{https://nanohub.org/groups/nemo_3d_distribution}.


%

\end{document}